\documentclass[aps,reprint,groupaddress,superscriptaddress,amssymb,prx]{revtex4-2}
\usepackage[utf8]{inputenc}
\usepackage{color}
\usepackage{xcolor}
\usepackage{graphicx}
\usepackage{stmaryrd}

\usepackage{mdframed}
\usepackage{physics}
\usepackage{amsmath}
\usepackage{amsfonts}
\usepackage{bm}
\usepackage[symbol]{footmisc}

\begin{document}
\preprint{APS/123-QED}

\title{Information propagation in Gaussian processes on multilayer networks }

\author{Giorgio Nicoletti}
\affiliation{ECHO Laboratory, École Polytechnique Fédérale de Lausanne, Lausanne, Switzerland}
\author{Daniel Maria Busiello}
\affiliation{Max Planck Institute for the Physics of Complex Systems, Dresden, Germany}

\begin{abstract}
\noindent Complex systems with multiple processes evolving on different temporal scales are naturally described by multilayer networks, where each layer represents a different timescale. In this work, we show how the multilayer structure shapes the generation and propagation of information between layers. We derive a general decomposition of the multilayer probability for continuous stochastic processes described by Fokker-Planck operators. In particular, we focus on Gaussian processes, for which this solution can be obtained analytically. By explicitly computing the mutual information between the layers, we derive the fundamental principles that govern how information is propagated by the topology of the multilayer network. In particular, we unravel how edges between nodes in different layers affect their functional couplings. We find that interactions from fast to slow layers alone do not generate information, leaving the layers statistically independent even if they affect their dynamical evolution. On the other hand, interactions from slow to fast nodes lead to non-zero mutual information, which can then be propagated along specific paths of interactions between layers. We employ our results to study the interplay between information and instability, identifying the critical layers that drive information when pushed to the edge of stability. Our work generalizes previous results obtained in the context of discrete stochastic processes, allowing us to understand how the multilayer nature of complex systems affects their functional structure.
\end{abstract}

\maketitle
    
\vspace{0.5cm}

\section{Introduction}
\noindent The dynamics of complex real-world systems is usually captured by multiple intertwined degrees of freedom that evolve on different spatial and temporal scales \cite{dedomenico2023more, schaffer2021mapping, honey2017switching, nicoletti2023emergence}. Chemical reaction networks \cite{radulescu2008robust, avanzini2023circuit, liang2022emergent}, neural systems \cite{timme2014multiplex, cavanagh2020diversity, mariani2022disentangling}, and ecological communities \cite{hastings2010timescales, poisot2015beyond, nicoletti2023emergent} provide only a few examples where such a multiscale nature naturally emerges. In particular, the presence of multiple timescales has been found to be a key driver of several emergent behaviors, such as the selection of chemical species out of equilibrium \cite{busiello2021dissipation, dass2021equilibrium, berton2020thermodynamics}, the large-scale propagation of epidemics in temporal networks \cite{valdano2015analytical, koher2019contact, cai2024epidemic}, and the quantification of dissipative features at the nanoscale \cite{busiello2020coarse, celani2012anomalous}. The most natural way to describe this class of systems is to employ the framework of multilayer networks \cite{dedomenico2016physics, bianconi2018multilayer, dedomenico2013mathematical, boccaletti2014structure, aleta2019multilayer}. A promising approach is that of associating a well-defined timescale to each layer and its nodes \cite{tewarie2021interlayer, nicoletti2024information}. These nodes may represent different degrees of freedom, such as conformational states of a molecule, or stochastic variables describing a dynamical evolution on the multilayer network. Edges between layers characterize the interactions between the different timescales and thus describe how they communicate with each other. Understanding how this inter-layer structure influences the evolution of the system as a whole is a fundamental question, whose answer might shed light on the pivotal drivers of emergent functionality in several real-world examples.

Information theory provides tools and concepts to understand, in very general terms, whether or not the presence of a physical interaction generates statistical dependencies \cite{cover1999elements}. In particular, recent works highlighted the relevance of the mutual information to pinpoint effective coupling induced by shared environment \cite{nicoletti2021mutual,nicoletti2022mutual}. Moreover, information-theoretic approaches are gaining momentum also in the context of biophysics \cite{tkavcik2016information,tkavcik2009optimizing,mattingly2021escherichia,sachdeva2021optimal,moor2023dynamic,bauer2023information,amano2022insights}, as they are proving to be useful to shed light on the functionality of biological and biochemical systems.

Furthermore, by employing information theory, recent work highlighted the connection between the topology of multilayer networks and their functional structure for discrete degrees of freedom in the presence of higher-order interaction \cite{nicoletti2024information}. In particular, it reveals under which conditions edges between nodes in different layers translate into effective couplings at the information level. These effective couplings can be understood in terms of a few principles describing how information is generated and propagated across timescales, ascribing different roles to inter-layer interactions according to their directionality. In this work, we extend this framework to the case of continuous stochastic processes on multilayer networks. Each node is characterized by a continuous real variable, and edges between the nodes are pairwise interactions between such degrees of freedom. We focus on the case of Gaussian processes, which provide solvable effective models in several biologically relevant cases \cite{di2024variance, busiello2024unraveling, dabelow2019irreversibility} and allow for analytical insights in different scenarios \cite{tostevin2009mutual, nicoletti2021mutual, nicoletti2022information}. Furthermore, they describe the linear response regime for non-linear stochastic dynamics close to their fixed points.

We introduce a simple graphical method to solve the multilayer dynamics, accounting for the separation of timescales between the layers and uncovering the full probabilistic structure of the system. From this structure, we study how information propagates across timescales and we recover the analogous principles of discrete systems with higher-order interactions. Interactions from slow to fast layers can generate information, while interactions from fast to slow layers can only propagate it downstream. We test our results both in fully-connected and sparse topologies, as well as in the presence of weighted and random adjacency matrices, and random interactions. In particular, our results in the timescale separation regime remain robust even in the case of small separation between the timescales. Finally, we show that the overall amount of information increases when the system is pushed toward the edge of instability. In particular, we find that this increase in information is driven by the slow regulating layers that generate information. Our work generalizes the results of \cite{nicoletti2024information} in a broader context and explores new features relevant to understanding the functional regime of operations of several complex systems.

\section{Methods}
\subsection{Gaussian processes on multilayer networks}
\noindent We consider a Gaussian process defined on a directed multilayer network with $N$ layers. Each layer $\mu = 1, \dots, N$ has $M_\mu$ nodes, connected among themselves by an adjacency matrix $\hat{A}_{\mu\mu}$. The structure of the multilayer network is described by the adjacency tensor
\begin{equation}
    \hat{A} = \begin{pmatrix}
        \hat{A}_{11} & \hat{A}_{12} & \cdots & \hat{A}_{1N} \\
        \hat{A}_{21} & \hat{A}_{22} & \cdots & \hat{A}_{2N} \\ 
        \vdots & \vdots & \ddots & \vdots \\
        \hat{A}_{N1} & \hat{A}_{N2} & \cdots & \hat{A}_{NN} \\ 
    \end{pmatrix}
\end{equation}
where the elements $A_{\mu\nu}^{ij}$ of the matrix $\hat{A}_{\mu\nu}$ are zero if and only if the $i$-th node of the $\mu$-th layer and the $j$-th node of the $\nu$-th layer are not connected. Otherwise, $A_{\mu\nu}^{ij}$ is the weight of the corresponding edge. We assume that $A_{\mu\mu}^{ii} = 0$. Notice that, for ease of notation in writing dynamical equations, we adopt the index convention that $A_{\mu\nu}^{ij}$ contains the directed connection from the $x_\nu^j$ to $x_\mu^i$.

We assume that the dynamical evolution is described by a Gaussian process $(\vec{x}_1(t), \dots, \vec{x}_N(t))$ on the nodes of all layers. Each layer $\mu$, in particular, evolves with a typical timescale $\tau_\mu$. The multilayer dynamics can be written as the Langevin equation
\begin{align}
\label{eqn:langevin_GP}
    \tau_\mu \dot{x}_\mu^i(t) = & - \left[x_\mu^i(t) - \theta_\mu^i\right] + \sum_{j = 1}^{M_\nu} A_{\mu\mu}^{ij}\left[x_\mu^j(t) - \theta_\mu^j\right] + \nonumber \\
    & + \sum_{\nu \ne \mu}\sum_{k = 1}^{M_\nu} A_{\mu\nu}^{ij}\left[x_\nu^k(t)-\theta_\nu^k\right] + \nonumber \\
    & + \sqrt{2 D_\mu^i} \, \xi_\mu^i(t) 
\end{align}
where $\tau_\mu$ is the timescale of the process in the $\mu$-th layer, $\theta_\mu^i$ is the mean of the process in the $i$-th node of the $\mu$-th layer, $D_\mu^i$ is its diffusion coefficient, and $\xi_\mu^i$ are independent white noises. In this case, the adjacency tensor represents interactions between the nodes, with each node representing a continuous Gaussian variable. The stability of this process is determined by the spectral radius of the tensor
\begin{equation}
\label{eqn:Wtensor}
    \hat{W} = \begin{pmatrix}
        \hat{A}_{11} - \mathbb{I}_{1} & \hat{A}_{12} & \cdots & \hat{A}_{1N} \\
        \hat{A}_{21} & \hat{A}_{22} - \mathbb{I}_2 & \cdots & \hat{A}_{2N} \\ 
        \vdots & \vdots & \ddots & \vdots \\
        \hat{A}_{N1} & \hat{A}_{N2} & \cdots & \hat{A}_{NN} - \mathbb{I}_N \\ 
    \end{pmatrix}
\end{equation}
where $\mathbb{I}_\mu$ is the $M_\mu \times M_\mu$ identity matrix. If the spectral radius $r$ of $\hat{W}$ is greater than zero, the Gaussian process is unstable and diverges in the long-time limit.

Eq.~\eqref{eqn:langevin_GP} can be cast as a Fokker-Planck equation for the multilayer probability $p_{1, \dots, N}(\vec{x}_1, \dots, \vec{x}_N, t)$ \cite{gardiner},
\begin{equation}
    \label{eqn:FP_GP}
    \frac{\partial}{\partial t} p_{1, \dots, N}(\vec{x}_1, \dots, \vec{x}_N, t) = \sum_{\mu = 1}^N\frac{1}{\tau_\mu}\mathcal{L}_\mu \,  p_{1, \dots, N}(\vec{x}_1, \dots, \vec{x}_N, t),
\end{equation}
where the differential operators $\mathcal{L}_\mu$ are given by
\begin{align}
    \mathcal{L}_\mu = - \nabla_\mu \cdot \biggl[& \hat{W}_{\mu\mu}\left(\vec{x}_\mu - \vec{\theta}_\mu\right) + \sum_{\nu \ne \mu}\hat{A}_{\mu\nu}\left(\vec{x}_\nu - \vec{\theta}_\nu\right) + \nonumber \\
    & - \hat{D}_\mu \nabla_\mu\biggl]
\end{align}
with $\hat{D}_\mu = \text{diag}(D_\mu^1, \dots, D_\mu^N)$. This Fokker-Planck equation can be solved analytically. In particular, its stationary distribution is given by the multivariate Gaussian distribution $\mathcal{N}(\cdot, \cdot)$:
\begin{equation}
    p^\mathrm{st}_{1, \dots, N}(\vec{x}_1, \dots, \vec{x}_N) = \mathcal{N}(\vec{\Theta}, \hat{\Sigma})
\end{equation}
where the mean is given by $\vec{\Theta} = (\vec{\theta}_1, \dots, \vec{\theta}_N)$ and the covariance matrix $\hat{\Sigma}$ obeys the Lyapunov equation
\begin{equation}
\label{eqn:Lyapunov_GP}
    (\hat{W} \odot \hat{T}) \hat{\Sigma} + \hat{\Sigma} (\hat{W} \odot \hat{T})^T = 2 \hat{D}
\end{equation}
where $\hat{T}$ is a matrix with $1/\tau_\mu$ in all entries of the $\mu$-th row, $\odot$ denotes the Hadamar or elementwise product, and $\hat{D}$ a diagonal tensor containing the diffusion matrices $\hat{D}_\mu$ of each layer,
\begin{equation}
    \hat{D} = \begin{pmatrix}
        \hat{D}_{1}/\tau_1 & \cdots & 0 \\
        \vdots & \ddots & \vdots \\
        0 & \cdots & \hat{D}_{N}/\tau_N \\ 
    \end{pmatrix} \; .
\end{equation}
As we can see from Eq.~\eqref{eqn:Lyapunov_GP}, the ratios between the different timescales of each layer enter explicitly in the covariance matrix, and thus we expect them to be instrumental in determining the dependencies between the layers.

\subsection{Mutual information for multilayer observables}
\noindent We are interested in quantifying effective couplings between different layers, i.e., the effective dependencies generated by the underlying multilayer topology. In general, at stationarity, we seek to evaluate the mutual information between layer $\mu$ and $\nu$, namely
\begin{equation}
    I_{\mu\nu} = \int d\vec{x}_\mu \, d\vec{x}_\nu \, p^\mathrm{st}_{\mu\nu}(\vec{x}_\mu, \vec{x}_\nu) \log_2 \frac{p^\mathrm{st}_{\mu\nu}(\vec{x}_\mu, \vec{x}_\nu)}{p^\mathrm{st}_{\mu}(\vec{x}_\mu) \, p^\mathrm{st}_{\nu}(\vec{x}_\nu)}
\end{equation}
where $p^\mathrm{st}_{\mu\nu}$, $p^\mathrm{st}_{\mu}$, and $p^\mathrm{st}_{\nu}$ are marginal probabilities, e.g.,
\begin{equation*}
    p^\mathrm{st}_{\mu\nu}(\vec{x}_\mu, \vec{x}_\nu) = \int \prod_{\alpha \ne \mu, \nu} d\vec{x}_\alpha \, p^\mathrm{st}_{1, \dots, N}(\vec{x}_1, \dots, \vec{x}_N) \; .
\end{equation*}
$I_{\mu\nu}$ is always non-negative and quantifies the dependency between the set of all nodes of two different layers, in terms of the Kullback-Leibler divergence between their joint probability distribution and the corresponding factorization \cite{cover1999elements}.

Since the multilayer probability is a Gaussian distribution, all marginal distributions will be Gaussians as well. Hence, the mutual information reads
\begin{equation}
    I_{\mu\nu} = \frac{1}{2}\log_2 \frac{\det\hat\Sigma_{\nu\nu}}{\det(\hat{\Sigma}_{\nu\nu} - \hat{\Sigma}_{\nu\mu}\hat\Sigma^{-1}_{\mu\mu}\hat\Sigma_{\mu\nu})}
\end{equation}
where the denominator is the determinant of the submatrix of $\hat\Sigma$ corresponding to the blocks of the $\mu$-th and the $\nu$-th layer, and $\hat{\Sigma}_{\mu\mu}$ and $\hat{\Sigma}_{\nu\nu}$ are the diagonal blocks of the covariance matrix. Similarly, we may write
\begin{equation}
    I_{\mu\nu} = \frac{1}{2}\log_2 \frac{\det\hat\Sigma_{\mu\mu}}{\det(\hat{\Sigma}_{\mu\mu} - \hat{\Sigma}_{\mu\nu}\hat\Sigma^{-1}_{\nu\nu}\hat\Sigma_{\nu\mu})}
\end{equation}
and the two expressions are equivalent. $I_{\mu\nu}$ defines a $N \times N$ symmetric matrix, named Mutual Information Matrix for Multiscale Observables (MIMMO) \cite{nicoletti2024information}. The MIMMO quantifies how much the layers are dependent on one another by measuring their shared information. In particular, as we will see, two layers may be densely connected but statistically independent due to the effect of their different timescales. In this particular case, although both $\hat{A}_{\mu\nu}$ and $\hat{A}_{\nu\mu}$ may be non-vanishing, we would find $I_{\mu\nu} = 0$. On the other hand, layers whose nodes are not directly connected by edges may become strongly coupled, implying that $I_{\mu\nu} > 0$.

\subsection{Timescale separation and conditional structure of the multilayer probability}
\noindent In principle, the mutual information between any two layers can be fully determined from $\hat\Sigma$. However, the analytical solution for the Lyapunov equation, Eq.~\eqref{eqn:Lyapunov_GP}, is in general cumbersome to analyze. Thus, we resort to a timescale separation limit \cite{bo2017multiple,busiello2020coarse,nicoletti2021mutual,nicoletti2024information} that allows us to derive the conditional structure of the multilayer probability. The main advantage of this approach is that such conditional structure does not require explicit knowledge of the dynamics, allowing us to draw general considerations valid for all stochastic processes in multilayer networks.

We now order the layers by their timescales, i.e., we take $\tau_1 < \tau_2 \dots < \tau_N$. We seek a solution of the Fokker-Planck equation for the multilayer probability distribution, Eq.~\eqref{eqn:FP_GP}, of the form
\begin{align}
\label{eqn:app:solution_form_ts}
    p_{1, \dots, N}(\vec{x}_1, \dots, \vec{x}_N, t) = \, p_{1, \dots, N}^{(1,0)}(\vec{x}_1, \dots, \vec{x}_N, t) + \\
    + \sum_{\mu = 2}^{N-2}\left(\prod_{\nu = 1}^{\mu -1}\epsilon_\nu\right)p_{1, \dots, N}^{(\mu, 0)}(\vec{x}_1, \dots, \vec{x}_N, t) + \nonumber \\
    + \prod_{\mu = 1}^N \epsilon_\mu \, p_{1, \dots, N}^{(N,1)}(\vec{x}_1, \dots, \vec{x}_N, t) + \mathcal{O}(\epsilon^2) \nonumber
\end{align}
where $\epsilon_\mu = \tau_\mu /\tau_N$, and $\mathcal{O}(\epsilon^2)$ contains all second-order contributions, for any $\mu = 1, \dots, N$. In Eq.~\eqref{eqn:app:solution_form_ts}, $p_{1, \dots, N}^{(\mu, 0)}$ denotes the zero-th order contribution in $\epsilon_\mu$, and $p_{1, \dots, N}^{(N, 1)}$ the first-order contribution in the slowest timescale, $\epsilon_N = 1$. This expression corresponds to subsequent expansions of the multilayer probability in $\epsilon_1, \dots, \epsilon_N$. In the limit $\epsilon_\mu \to 0$, the solution at the leading order, $p_{1, \dots, N}^{(1,0)}$, obeys
\begin{align}
\label{eqn:leading_order_me}
    \partial_t p_{1, \dots, N}^{(1,0)}(\vec{x}_1, \dots, \vec{x}_N, t) &= \mathcal{L}_N\, p_{1, \dots, N}^{(1,0)}(\vec{x}_1, \dots, \vec{x}_N, t) \nonumber \\
    &+ \sum_{\mu = 1}^{N-1} \frac{\mathcal{L}_\mu}{\epsilon_\mu} \, p_{1, \dots, N}^{(1,0)}(\vec{x}_1, \dots, \vec{x}_N, t) \nonumber \\
    &+ \mathcal{L}_1 p_{1, \dots, N}^{(2,0)}(\vec{x}_1, \dots, \vec{x}_N, t)  
\end{align}
after rescaling time by the slowest timescale.

We solve Eq.~\eqref{eqn:leading_order_me} order-by-order, with each order corresponding to a given layer. The first order is associated with the layer with the fastest dynamics - the first layer, in our ordering - and is immediately solved by the stationary solution of
\begin{equation*}
    0 = \mathcal{L}_1\left(\{x\}_{\rightsquigarrow 1}\right)\,p^\mathrm{st}_{1 | \rho(1)}(\vec{x}_1 | \{\vec{x}_\nu\}_{\nu \in \rho(1)})
\end{equation*}
where $\{x\}_{\rightsquigarrow 1} \equiv \rho(1)$ both denote the set of all slower layers for which at least one node is connected to a node of the first layer. Notice that, in this solution, all layers that interact with the first one enter as conditional dependencies, which amounts to solving the first layer's Fokker-Planck equation while keeping the value of all nodes in all other layers fixed - i.e., by freezing their dynamics. Furthermore, the first order solution implies that $p_{1, \dots, N}^{(1,0)} = p^\mathrm{st}_{1 | \rho(1)}\,p_{2, \dots, N}^{(1,0)}$.

The second order, after an integration over $\vec{x}_1$, is the stationary solution of the effective operator
\begin{equation*}
    \mathcal{L}^\mathrm{eff}_{2 | \rho(2)} = \int d\vec{x}_1 \mathcal{L}_2\left(\{x\}_{\rightsquigarrow 2}\right) \, p^\mathrm{st}_{1 | \rho(1)}(\vec{x}_1 | \{\vec{x}_\nu\}_{\nu \in \rho(1)})
\end{equation*}
so that
\begin{equation*}
    0 = \mathcal{L}^\mathrm{eff}_{2 | \rho(2)} p^{\mathrm{eff, st}}_{2 | \rho(2)} (\vec{x}_2 | \{\vec{x}_\nu\}_{\nu \in \rho(2)}) \;.
\end{equation*}
Here, $\rho(2)$ denotes all conditional dependencies that remain in the effective Fokker-Planck operator $\mathcal{L}^\mathrm{eff}_{2 | \rho(2)} \equiv \mathcal{L}^\mathrm{eff}_{2 | \rho(2)}\left(\vec{x}_2 | \{\vec{x}_\nu\}_{\nu \in \rho(2)}\right)$ after the integration over $\vec{x}_1$. In particular, the stationary probability $p^{\mathrm{eff, st}}_{2 | \rho(2)}$ will depend on all slower layers connected to the second. Furthermore, $\rho(2)$ may also inherit dependencies on slower layers from all faster layers that have been integrated out - in this case, only from $\rho(1)$ appearing in $p^\mathrm{st}_{1 | \rho(1)}$. Indeed, if the first layer is connected to the second, the conditional dependencies contained in $\rho(2)$ will include all layers directly connected to the first and slower than the second. Otherwise, without interactions from the first to the second layer, the effective operator will not inherit dependencies from $\rho(1)$.

By recursively solving each order and marginalizing over the faster layers, we obtain a hierarchy of effective Fokker-Planck operators:
\begin{equation}
\label{eqn:effective_operators}
    \mathcal{L}^\mathrm{eff}_{\mu | \rho(\mu)} = \int d\vec{x}_1 \dots d \vec{x}_{\mu -1} \mathcal{L}_\mu\left(\{x\}_{\rightsquigarrow \mu}\right) \prod_{\nu = 1}^{\mu - 1} p_{\nu | \rho(\nu)}^{\rm eff, st}
\end{equation}
where $p_{\nu | \rho(\nu)}^{\rm eff, st} (\vec{x}_\nu | \{\vec{x}_\alpha\}_{\alpha \in \rho(\nu)})$ is the stationary probability of the $\nu$-th layer, obeying
\begin{equation}
    \mathcal{L}^\mathrm{eff}_{\nu | \rho(\nu)} \, p_{\nu | \rho(\nu)}^{\rm eff, st} (\vec{x}_\nu | \{\vec{x}_\alpha\}_{\alpha \in \rho(\nu)}) = 0 \; .
\end{equation}
Thus, in general, we can always solve the multilayer Gaussian process recursively, starting from the fastest layer. The multilayer probability distribution in this timescale separation regime, at the leading order, is then $p^\mathrm{eff}_{1, \dots, N} \equiv p^{(1,0)}_{1, \dots, N}$ and reads
\begin{equation}
\label{eqn:multilayer_probability}
    p^\mathrm{eff}_{1, \dots, N}(t) = p_N^\mathrm{eff}(\vec{x}_N, t) \prod_{\nu = 1}^{N-1}{\color{black}p_{\nu | \rho(\nu)}^{\rm eff, st}}(\vec{x}_\nu | \{\vec{x}_\alpha\}_{\alpha \in \rho(\alpha)})
\end{equation}
which makes the conditional structure manifest. Since all $p_{\nu | \rho(\nu)}^{\rm eff, st}$ remain Gaussian, by definition, also $p^\mathrm{eff, st}_{1, \dots, N}$ is Gaussian as well. However, this factorization will allow us to infer the general features of how information is generated and propagated across layers.

In particular, by explicitly carrying out the recursive solution, it is possible to interpret the set $\rho(\mu)$ in terms of the topology of the multilayer network. Consider the directed graph $\mathcal{G}$ with $N$ nodes associated with the multilayer network, where each node now represents a layer, and edges are given by interactions between layers. In this coarse-grained graph, we name interactions going from a fast layer to a slow layer ``direct interactions'', i.e., edges contained in $\hat{A}_{\mu\nu}$ with $\mu > \nu$. On the other hand, interactions going from a slow to a fast layer are ``feedback interactions'', with $\mu < \nu$. Following \cite{nicoletti2024information}, we define a propagation path (PP) on $\mathcal{G}$ as a directed path from layer $\mu$ to layer $\nu$ with $\mu > \nu$ and such that it contains at least one direct interaction, i.e., one edge $\alpha \to \beta$ with $\alpha < \beta$. Then, we consider the graph $\mathcal{G}^{(\nu)}(\nu^*)$ as the induced subgraph obtained by removing all layers slower than $\nu$ - i.e., all nodes $\alpha > \nu$ - except $\nu^*$. A propagation path from $\mu$ to $\nu$ is a minimal propagation path if it is a PP both in $\mathcal{G}$ and in the induced subgraph $\mathcal{G}^{(\nu)}(\mu)$. Then, we define $\rho(\mu)$ as the set
\begin{equation}
\label{eqn:rho}
    \rho(\mu) = \left\{ \nu >\mu : \exists \text{ mPP } \nu \to \mu \text{ or } \hat{C}_{\nu\mu} \ne 0\right\}
\end{equation}
which contains all layers connected to $\mu$ either via an mPP or a single feedback interaction. Thus, by inspecting interactions between layers in the coarse-grained graph $\mathcal{G}$, we are able to identify a rule to build $\rho(\mu)$ only from the topology of edges between layers.

\section{Results}
\subsection{Graphical construction of the solution}
\label{sec:graphical}
\noindent Since we are dealing with Gaussian processes, we have shown that the multilayer stationary probability has the form of a product of Gaussian distributions (see Eq.~\eqref{eqn:multilayer_probability}). Here, we show how the mean and variance of these distributions can be directly constructed from the topology of inter-layer connections with a graphical method.

We start with the simplest case of a two-layer system with a direct interaction from the first to the second layer. The adjacency tensor is:
\begin{equation}
    \hat{A} = \begin{pmatrix}
        \hat{A}_{11} & 0 \\
        \hat{A}_{21} & \hat{A}_{22}
    \end{pmatrix}
    \label{ADJ}
\end{equation}
so that $\hat{A}_{21}$ contains the edges from layer $1$ to layer $2$. The dynamics of this system is given by:
\begin{eqnarray*}
    \tau_1 x^i_1(t) &=& \sum_j W^{ij}_{11} \left(x^j_1(t) - \theta^j_1\right) + \sqrt{2 D^i_1} \xi^i_1(t) \\
    \tau_2 x^i_2(t) &=& \sum_j W^j_{22} \left( x^j_2(t) - \theta^j_2 \right) \\
    &+& \sum_j A^{ij}_{21} \left( x^j_1(t) - \theta^j_1 \right) + \sqrt{2 D^i_2} \xi^i_2(t)
\end{eqnarray*}
where $W_{\mu\mu}^{ij} = A_{\mu\mu}^{ij} - \delta^{ij}_{\mu\mu} $ and $\xi^i_\mu$ are independent white noises. At stationarity, according to the timescale separation solution, the joint probability is the product of two Gaussian distributions. In Appendix \ref{app:solution_gaussian}, by carrying on the calculations explicitly, we show that:
\begin{equation}
    p^{\rm st}_{1,2}(\vec{x}_1, \vec{x}_2) = \mathcal{N}_1(\vec{\theta}_1, \hat{S}_{1}) \mathcal{N}_2(\vec{\theta}_2, \hat{S}_{2}) \;.
\end{equation}
Here, $\mathcal{N}_\mu$ denotes a Gaussian distribution for $\vec{x}_\mu$, while $\hat{S}_\mu$ is the covariance matrix of the layer $\mu$ obtained if it were independent, i.e., the one satisfying the Lyapunov equation:
\begin{equation}
    \hat{W}_{\mu\mu} \hat{S}_\mu + \hat{S}_\mu \hat{W}_{\mu\mu}^T = 2 \hat{D}_{\mu} \;.
\end{equation}
Thus, when the faster layer regulates the slower one, the multilayer probability explicitly decomposes in the product of the two single-layer probabilities as they were uncoupled, in the timescale separation regime.

Consider now a two-layer system where the interaction goes from the second to the first layer, i.e., a feedback interaction. In this scenario, $\hat{A}_{21} = 0$ and the adjacency matrix will contain a non-zero off-diagonal block $\hat{A}_{12}$. Following Appendix \ref{app:solution_gaussian}, the explicit solution in timescale separation can be obtained and is given by:
\begin{equation}
    p^{\rm st}_{1,2}(\vec{x}_1, \vec{x}_2) = \mathcal{N}_1(\vec{m}_1(\vec{x}_2), \hat{S}_{1}) \mathcal{N}_2(\vec{\theta}_2, \hat{S}_{2}) \;,
\end{equation}
where $\vec{m}_1(\vec{x}_2) = \vec\theta_1 - \hat{W}_{11}^{-1} \hat{A}_{12} \left( \vec{x}_2 - \vec{\theta}_2 \right)$. In this case, the second layer is unaffected by the first, and its stationary distribution coincides with the independent case. However, the mean of $\mathcal{N}_1$ is regulated by the second layer and explicitly depends on $\vec{x}_2$.

Before building a general graphical rule, we study what happens in the presence of a minimal propagation path. Consider the following adjacency tensor:
\begin{equation*}
    \hat{A} = \begin{pmatrix}
        \hat{A}_{11} & 0 & \hat{A}_{13} \\
        \hat{A}_{21} & \hat{A}_{22} & 0 \\
        0 & 0 & \hat{A}_{33} \;.
    \end{pmatrix}
\end{equation*}
Here, the minimal propagation path (mPP) is $\vec{x}_3 \to \vec{x}_1 \to \vec{x}_2$, i.e., there is a feedback interaction from $3$ to $1$ and a direct from $1$ to $2$. The explicit expression for the multilayer probability, derived in Appendix \ref{app:solution_propagation}, is:
\begin{equation}
    p^{\rm st}_{1,2,3} = \mathcal{N}_1(\vec{m}_1(\vec{x}_3), \hat{S}_{1}) \mathcal{N}_2(\vec{m}_2(\vec{x}_3), \hat{S}_{2}) \mathcal{N}_3(\vec{\theta}_3, \hat{S}_{3}) \;,
\end{equation}
where $\vec{m}_1(\vec{x}_3) = \vec\theta_1 - \hat{W}_{11}^{-1} \hat{A}_{13} (\vec{x}_3 - \vec{\theta}_3)$ and $\vec{m}_2(\vec{x}_3) = \vec\theta_2 - \hat{W}_{22}^{-1} \hat{A}_{21} \hat{W}_{11}^{-1} \hat{A}_{13} (\vec{x}_3 - \vec{\theta}_3)$. As before, the mean of $\mathcal{N}_1$ is modified due to the feedback interaction from the third layer. However, the mean of $\mathcal{N}_2$ also retains the effect of the mPP. This results in a concatenation of similar building blocks, backtracing the mPP from layer $2$ to its origin, layer $3$.

Overall, in the timescale separation limit, only feedback interactions and mPPs modify the stationary distribution in a way that can be understood from the topology of the multilayer network. In particular, introducing the elementary building block $\hat\Omega_{ji} = \hat{W}_{jj}^{-1} \hat{A}_{ji}$, the steady-state distribution for a given layer $\mu$ is:
\begin{equation*}
    \mathcal{N}_\mu \left( \vec\theta_\mu - \sum_{\nu \in \rho(\mu)} \Bigg( \prod_{(\kappa, \lambda) \in \{ \mu \Rightarrow \nu \}} \hat\Omega_{\kappa \lambda} \Bigg) (\vec{x}_\nu - \vec\theta_\nu), \hat{S}_\mu \right) \;,
\end{equation*}
where $\{ \mu \Rightarrow \nu \}$ is a short-hand notation to indicate the set of all edges that connect $\nu$ to $\mu$ through a single feedback (i.e., one single edge) or a mPP. The direction of the arrow indicates the order in which all these edges have to be taken, i.e., from $\mu$ to $\nu$.

\subsection{Feedback and direct interactions}
\begin{figure}
    \centering
    \includegraphics[width = \columnwidth]{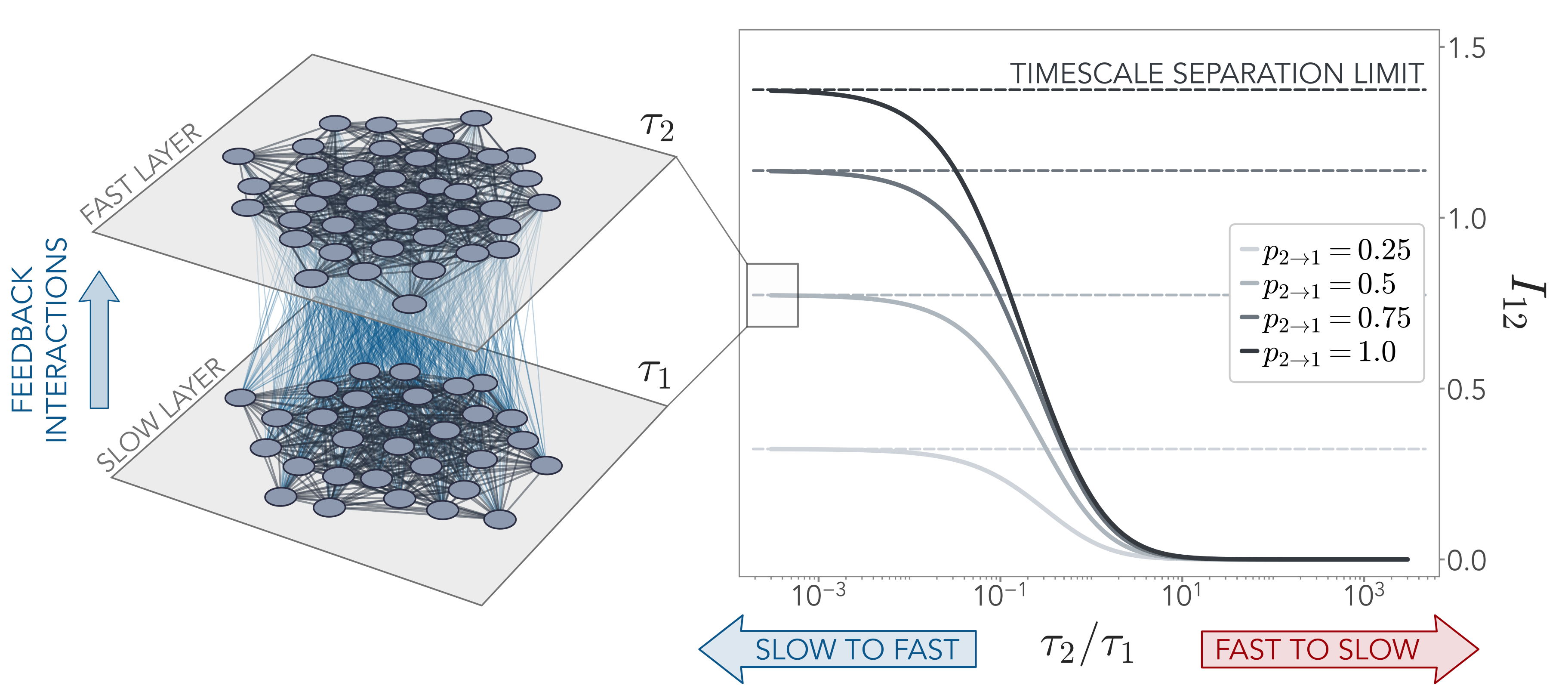}
    \caption{Mutual information between layers in a two-layer network, with directed interactions going from layer one, evolving with a timescale $\tau_1$, to layer two, evolving with a timescale $\tau_2$. Each layer is an Erdos-Renyi network with $p_\mathrm{conn} = 0.5$. The interaction matrix $\hat{A}_{21}$ may be sparse as well. If $\tau_1 \ll \tau_2$, we have ``direct interactions'' from a fast to a slow layer, and no mutual information is present. In the opposite limit, we have ``feedback interactions'' from a slow to a fast layer, which generate information in the system, so that $I_{12} > 0$. A denser interaction matrix leads to higher $I_{12}$. The results are averaged over $10^3$ realizations of the multilayer network, with $100$ nodes in layer one, and $150$ nodes in layer two.}
    \label{fig:figure1}
\end{figure}

\noindent To understand the effect of timescales on the shared information between Gaussian processes on different layers, we first consider again the simple case of a two-layer system. Each layer is an Erdos-Renyi network with a different number of nodes $M_1$ and $M_2$ (see Figure \ref{fig:figure1}) and its adjacency tensor is the one in Eq.~\eqref{ADJ}. We can think of this example as the first layer regulating the dynamics of the second. Since we are interested in exploring all possible timescale regimes, we are not making any assumption on $\tau_1/\tau_2$ for now.
We set $\hat{A}_{21}$ to be a sparse matrix, with a probability of connection $p_{1 \to 2}$, and elements $A_{21}^{ij} = 1/M_1$ if $j$ in the first layer interacts with $i$ in the second. The dynamics of this system is given by
\begin{equation*}
\begin{gathered}
    \tau_1 \vec{x}_1(t) = \hat{W}_{11} \vec{x}_1(t) + \sqrt{2} \vec{\xi}_1(t) \\
    \tau_2 \vec{x}_2(t) = \hat{W}_{22} \vec{x}_2(t) + \hat{A}_{21}\vec{x}_1(t) + \sqrt{2} \vec{\xi}_2(t)
\end{gathered}
\end{equation*}
where $W_{\mu\mu}^{ij} = A_{\mu\mu}^{ij} - \delta^{ij}_{\mu\mu} $, $\vec{\xi}_\mu$ are independent white noises, and for simplicity we are considering Gaussian processes with zero mean and unitary diffusion coefficient. In the previous section, we have shown that it is always possible to generalize the results to processes with a non-zero mean and a non-unitary diffusion coefficient.

\begin{figure*}
    \centering
    \includegraphics[width = \textwidth]{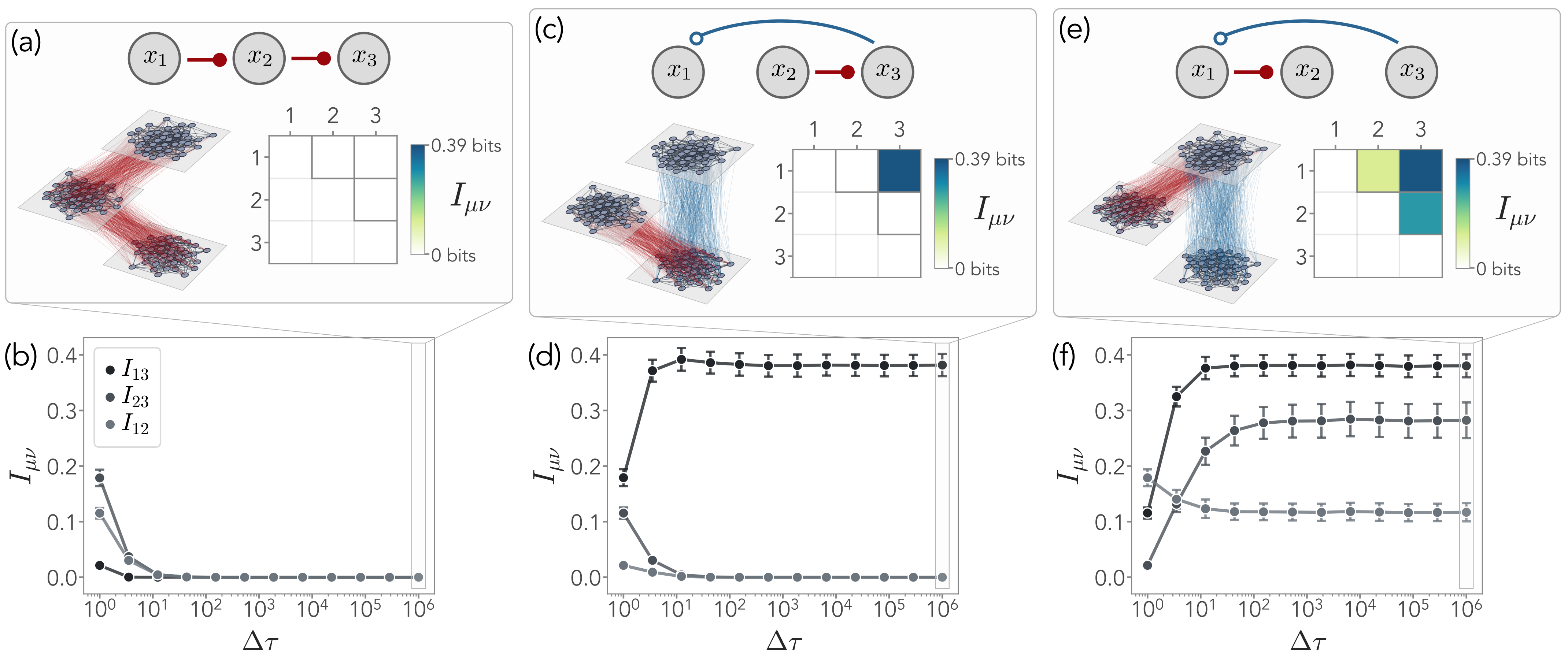}
    \caption{The principles of information propagation in multilayer networks. Each layer is a fully connected network with a Gaussian adjacency matrix $A_{\mu\mu}^{ij} \sim \mathcal{N}(0, \sigma^2/M_\mu)$, with $\sigma = 0.5$, and results are averaged over $10^4$ realizations. The timescales of the layers are separated by $\tau_{\mu+1}/\tau_{\mu} = \Delta \tau$. (a-b) A three-layer network with interactions $\vec{x}_1 \to \vec{x}_2 \to \vec{x}_3$. At large $\Delta \tau$ these are direct interactions, and all the elements of the MIMMO are zero. The timescale separation qualitatively works already at $\Delta \tau \approx 10^{1}$. (c-d) A three-layer network with interactions $\vec{x}_2 \to \vec{x}_3 \to \vec{x}_1$. In the timescale separation regime, the feedback interaction from layer $3$ acts as an information source for layer $1$, so that $I_{13} \neq 0$ in the MIMMO. Once more, the isolated direct link does not generate information. (e-f) A three-layer network with interactions $\vec{x}_3 \to \vec{x}_1 \to \vec{x}_2$. All the elements of the MIMMO are non-zero since $3 \to 1 \to 2$ is a minimal propagation path. The feedback link from $3$ to $1$ generates information, which is then propagated to $2$ by the direct link.}
    \label{fig:figure2}
\end{figure*}

At stationarity, the joint probability between the layers, $p^\mathrm{st}_{12}(\vec{x}_1, \vec{x}_2)$ is a Gaussian probability with zero mean and covariance matrix $\hat{\Sigma}$ obeying the Lyapunov equation
\begin{equation*}
    \begin{pmatrix}
        \frac{\hat{W}_{11}}{\tau_1} & 0 \\
        \frac{\hat{A}_{21}}{\tau_2} & \frac{\hat{W}_{22}}{\tau_2}
    \end{pmatrix}
    \hat{\Sigma} + \hat{\Sigma}
    \begin{pmatrix}
        \frac{\hat{W}_{11}^T}{\tau_1} & \frac{\hat{A}_{21}}{\tau_2} \\
        0 & \frac{\hat{W}_{22}^T}{\tau_2}
    \end{pmatrix}
    = 2
    \begin{pmatrix}
        1/\tau_1 & 0 \\
        0 & 1/\tau_2
    \end{pmatrix}
\end{equation*}
where we carried out explicitly the Hadamar product in Eq.~\eqref{eqn:Lyapunov_GP} (see Appendix \ref{app:solution_gaussian}). In Figure \ref{fig:figure1}, we show the behavior of $I_{12}$ as a function of the relative timescale separation between the two layers, $\tau_1/\tau_2$. We find that no information is shared between the layers when $\tau_1 \ll \tau_2$, i.e., when the regulating layer - the first one - evolves faster than the regulated layer - the second one. In this case, the interaction is of the direct type. This can be immediately understood from the timescale separation solution, which is
\begin{equation*}
    p^\mathrm{eff, st}_{12}(\vec{x}_1, \vec{x}_2) = p^{\mathrm{st}}_1(\vec{x}_1) p_2^\mathrm{eff, st}(\vec{x_2}) = \mathcal{N}_1(0, \hat S_{1}) \, \mathcal{N}_2(0, \hat S_{2})
\end{equation*}
directly from the graphical solution in Sec.~\ref{sec:graphical}.

However, information increases and saturates when the dynamics of $\vec{x}_1$ becomes slower than that of $\vec{x}_2$, i.e., when $\tau_1 \gg \tau_2$. In this case, the interaction is of the feedback type, and the timescale separation solution becomes
\begin{equation*}
    p^\mathrm{eff, st}_{12} = p^\mathrm{st}_{1} p^\mathrm{st}_{2|1} = \mathcal{N}_1(0, \hat{S}_{1}) \, \mathcal{N}_2(\hat{\Omega}_{21} \vec{x}_1, \hat{S}_{2}) \;.
\end{equation*}
The conditional dependency on $p^\mathrm{st}_{2|1}$ now makes the two layers statistically dependent, generating information between them. Notably, the sparsity of the interactions between the two layers affects information, with sparser networks generating less information (Figure \ref{fig:figure1}). This simple example suggests a prime role of the directionality of interactions with respect to the corresponding timescales, as previously found for master equation dynamics and higher-order structures \cite{nicoletti2024information}.

\subsection{Information propagation across layers}
\noindent We now outline the general principles of information propagation on multilayer networks by focusing on a few simple examples. Let us consider the simple three-layer networks in Figure \ref{fig:figure2}. Once more, we order the layers by their timescale, and assume that the timescales are separated by $\Delta \tau > 1$, so that $\tau_{\mu + 1} / \tau_\mu = \Delta \tau$. To exploit the generality of our framework, we further take the adjacency matrix of each layer to be a random matrix \cite{may1972will,allesina2012stability,busiello2017entropy,luo2006application}, with elements following a Gaussian distribution:
\begin{equation}
    A_{\mu\mu}^{ij} \sim \mathcal{N}\left(0, \frac{\sigma^2}{M_\mu}\right)
\end{equation}
where $\sigma^2$ is the variance of the entries, and the scaling with $M_\mu$ ensures that the results do not depend on the number of nodes \cite{tao2008random}. Without loss of generality, we take once more interactions between layers to be $A_{\mu\nu}^{ij} = 1/M_\nu$ if there is an edge going from $x_{\nu}^j$ to $x_{\mu}^i$.

We begin with a simple multilayer network where the first layer influences the second, and the second influences the third (Figure \ref{fig:figure2}a). In this case, all interactions are direct interactions. From Eq.~\eqref{eqn:multilayer_probability} and Eq.~\eqref{eqn:rho}, it is immediately clear that the multilayer probability in this case is exactly factorizable, and thus all elements of the MIMMO are zero. Thus, despite the interactions, the timescale separation between the layers makes them statistically independent. In Figure \ref{fig:figure2}b, we show that this result remains qualitatively valid even below $\Delta \tau \approx 10^1$, suggesting that timescales affect information already at small relative separation. Thus, from this example and from the structure in Eq.~\eqref{eqn:multilayer_probability}, we find the first principle of information propagation in multilayer networks: (i) direct interactions alone do not generate information.

We now change the topology of interactions between layers (Figure \ref{fig:figure2}c). We consider a system where the second layer is connected to the third one, which in turn interacts with the first one. From the definition of minimal propagation paths, it is immediate to check that this is not an mPP. The timescale separation solution reads:
\begin{equation*}
    p^\mathrm{eff, st}_{123}(\vec{x}_1, \vec{x}_2, \vec{x}_3) = \mathcal{N}_1(\hat{\Omega}_{13} \vec{x}_3, \hat{S}_1) \, \mathcal{N}_2(0, \hat{S}_2) \, \mathcal{N}_3(0, \hat{S}_3) \;,
\end{equation*}
following the graphical solution introduced above. Thus, as in the previous examples, the direct interactions do not couple $\vec{x}_2$ and $\vec{x}_3$ at the information level, whereas the feedback interaction generates information from $\vec{x}_3$ to $\vec{x}_1$. Once more, the MIMMO in the timescale-separated regimes qualitatively captures the behavior of the system even at relatively similar timescales, emphasizing the robustness of our results beyond the limit in which they have been derived (Figure \ref{fig:figure2}d). This allows us to derive the second principle of information propagation in multilayer networks, which is manifest in Eq.~\eqref{eqn:rho}: (ii) feedback interactions generate information from slow to fast layers.

Finally, we consider the topology depicted in Figure \ref{fig:figure2}e. The path $\vec{x}_3 \to \vec{x}_1 \to \vec{x}_1$ is now an mPP, leading to the solution:
\begin{align*}
    p^\mathrm{eff, st}_{123}(\vec{x}_1, \vec{x}_2, \vec{x}_3) = & ~\mathcal{N}_1(\hat{\Omega}_{13} \vec{x}_3, \hat{S}_1) \, \mathcal{N}_2(\hat{\Omega}_{21} \hat{\Omega}_{13} \vec{x}_3, \hat{S}_2) \times \\
    & \times \mathcal{N}_3(0, \hat{S}_3) \;.
\end{align*}
Thus, it is clear that no pair of layers is independent anymore, because the second layer inherits a conditional dependence on the third layer through the first. In terms of the MIMMO, the information generated from layer $3$ to $1$ through the feedback interaction is then propagated to layer $2$ through the direct interaction. In Figure \ref{fig:figure2}d we show how the elements of the MIMMO change with $\Delta \tau$. This suggests one last principle, which follows from the definition of mPPs: (iii) information generated through feedback by a slow layer may be propagated through direct interactions to any faster layer.

Hence, the topological structure of the interactions between the timescale-ordered layers determines the propagation of information among them. The first two principles we uncover highlight the interplay between direct and feedback interactions and the information between layers. The third one leads to a natural definition of directionality in information propagation in terms of the timescales of the dynamics in the different layers.

\begin{figure}
    \centering
    \includegraphics[width = \columnwidth]{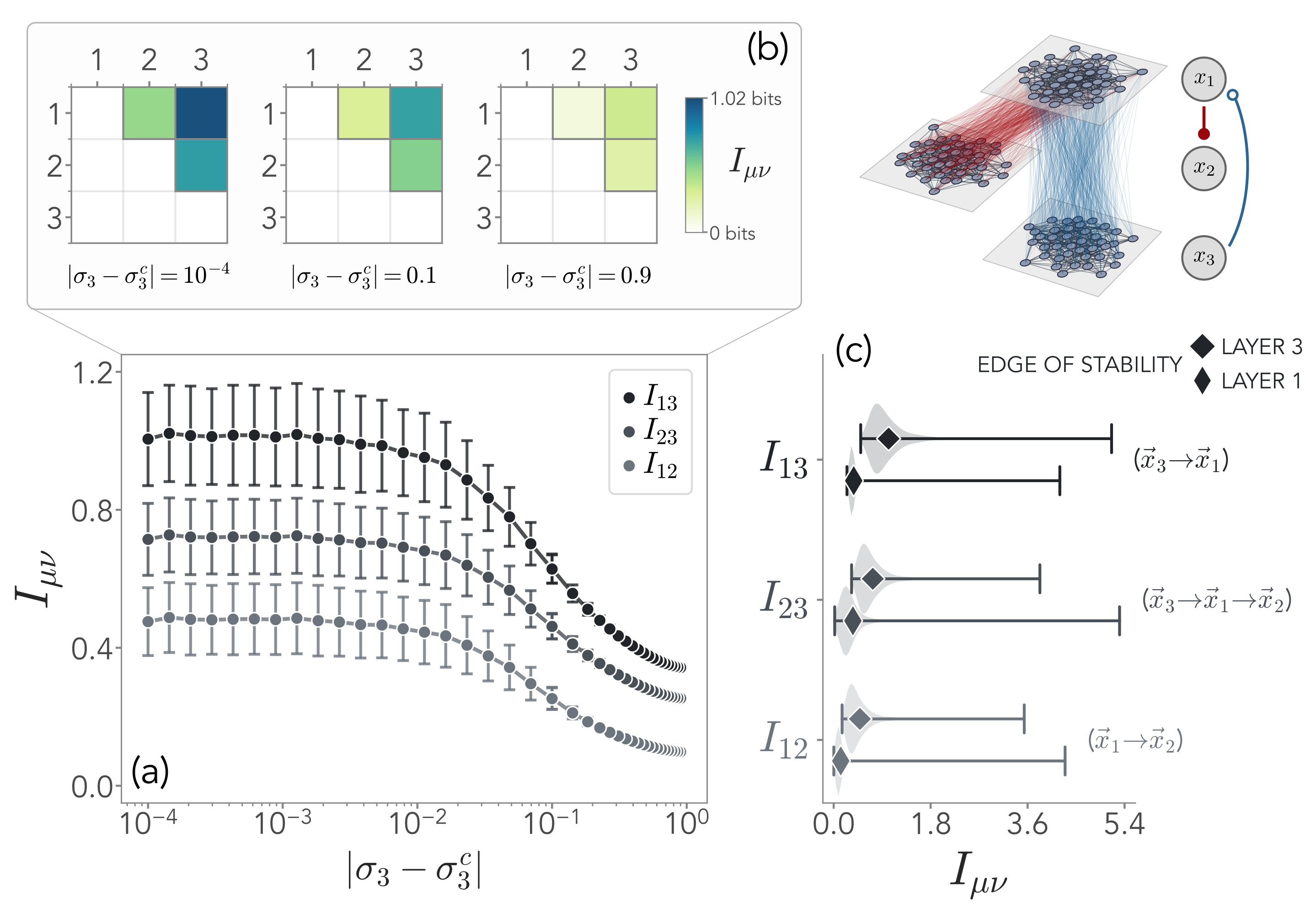}
    \caption{The timescales and the stability of the Gaussian processes in each layer affect information. Each layer is a fully connected network with a Gaussian adjacency matrix $A_{\mu\mu}^{ij} \sim \mathcal{N}(0, \sigma_\mu^2/M_\mu)$, with the edge of stability being $\sigma_\mu^c = 1$. Results are averaged over $10^4$ realizations of the edge weights. (a-b) In a three-layer network with interactions $\vec{x}_3 \to \vec{x}_1 \to \vec{x}_2$ all elements of the MIMMO are different from zero, and they all increase as $\sigma_3$ approaches the edge of stability $\sigma_3^c$ ($\sigma_1 = \sigma_2 = 0.1$). Error bars represent $1/3$rd of the standard deviation. (c) The information increase across all layers is on average larger if layer $3$ is close to instability. The shaded area represents the distribution over $10^4$ realizations. This can be understood in terms of information propagation, as information is generated from the slow layer $3$ to the fast layer $1$. If instead layer $1$ approaches the edge of stability, the average information increase is smaller on average. However, the variance of the elements of the MIMMO involving the direct interactions $\vec{x}_1 \to \vec{x}_2$, i.e., $I_{12}$ and $I_{23}$, is higher.}
    \label{fig:figure3}
\end{figure}

\subsection{Information propagation at the edge of stability}
\noindent We now exploit the fact that our framework can be applied to multilayer networks with random edge weights to study how information propagation is affected by the stability of each layer. To this end, we assume that the entries of the adjacency matrix of each layer follow the Gaussian distribution:
\begin{equation}
    A_{\mu\mu}^{ij} \sim \mathcal{N}\left(0, \frac{\sigma_\mu^2}{M_\mu}\right)
\end{equation}
where $\sigma_\mu^2/M_\mu$ is now a layer-dependent variance. For a Gaussian process of the form of Eq.~\eqref{eqn:langevin_GP}, a layer becomes unstable at a critical variance $\sigma_\mu^c = 1$ \cite{tao2008random}, making the entire system unstable as a consequence. We focus on the simple mPP in Figure \ref{fig:figure2}e, where the interactions between the three layers of the network follow $\vec{x}_3 \to \vec{x}_1 \to \vec{x}_2$. From the principles outlined in the previous section, one can heuristically understand the MIMMO in this case as information being generated from the third to the first layer, which is then propagated to the second. This suggests that the dynamics on the third layer acts as the source of information in the system.

In Figure \ref{fig:figure3}a, we show that the mutual information between all layers increases as $|\sigma_3 - \sigma_3^c| \to 0$, i.e., when the third layer approaches the edge of stability, while the other two are kept stable at $\sigma_1 = \sigma_2 = 0.1$. Notably, the variance of $I_{\mu\nu}$ over the realization of the edge weights sharply increases as well. Yet, the structure of the MIMMO remains the same at all values of $\sigma_3$, as well as the hierarchy among its entries ($I_{13} > I_{23} > I_{12}$), showing the robustness of the principles of information propagation even for dynamics close to instability (Figure \ref{fig:figure3}b). This emphasizes that the pivotal role in determining information propagation is played by the relative timescales between layers.

These results are qualitatively different when the unstable dynamics becomes that of the first layer, $\vec{x}_1$. We compare the two cases in Figure \ref{fig:figure3}c. With respect to the previous case, the average mutual information between all pairs of layers is lower when the first layer is at the edge of instability. However, both $I_{12}$ and $I_{23}$ display a larger variance. This suggests that when direct interactions propagate information through an unstable layer, the MIMMO will remain small on average, but the specific realization of the edge weights may allow for strong fluctuations. Importantly, this is not the case for $I_{13}$, since the source of the information comes from a stable Gaussian process. This further highlights how the underlying directionality of interactions plays a prime role in allowing us to interpret the functional couplings between the layers in terms of mutual information.

\section{Discussions}
\noindent In this work, we have characterized how timescales shape the features of Gaussian processes on multilayer networks. By ordering the layers according to their timescale and considering a timescale separation regime, we have derived a general decomposition of the joint probability that makes the conditional dependencies between the layers manifest. From this joint probability, we have shown that the topology of interactions across layers determines how information among them is generated and propagated. Our results are qualitatively robust even at small separations between the timescales, and generalize previous results obtained for discrete stochastic processes \cite{nicoletti2024information}. In particular, in a timescale-ordered multilayer network, the directionality of the edges between the layers is pivotal. Direct interactions from fast to slow layers do not result in any statistical dependency and thus information. However, information is generated from feedback interactions stemming from slow layers, and can then be propagated through minimal propagation paths. Finally, we have studied the interplay between information and stability, allowing us to understand which are the critical layers that lead to a maximal increase of information when pushed to the edge of stability.

Extensions of this work are manifold. In particular, we expect Eq.~\eqref{eqn:multilayer_probability} to be valid for any Fokker-Planck operator. The main advantage of considering Gaussian processes is that they are exactly solvable, and thus they allowed us to verify the robustness of our solution away from the timescale separation limit. However, it will be paramount to study how information is affected by more complex dynamics, as non-linearities have been shown to deeply affect information \cite{nicoletti2022mutual}. In particular, in the case of Gaussian processes, we have seen that sparsity in the interactions between the layers simply decreases information (Figure \ref{fig:figure1}). We expect this result to be significantly different in more general scenarios, where the topology of the adjacency tensor will be crucial to determining the dependencies between the layers. This may be particularly relevant in, e.g., the study of recurrent neural networks (RNNs), where non-linear activation functions integrate information from all incoming connections, and timescales are known to be crucial in determining learning and performance \cite{hihi1995hierarchical, chung2016hierarchical, wang2020two}. Similar considerations may be drawn for brain activity, where vastly different timescales are at play \cite{timme2014multiplex, wang2019hierarchical, cavanagh2020diversity, zeraati2023intrinsic}. Finally, modulation from slow timescales has been shown to produce critical-like dynamical features \cite{nicoletti2020scaling, morrell2021latent, mariani2022disentangling, morrell2024neural}, and in general both the topology of interactions and the timescales at play are crucial to understanding the behavior of models of neural activity \cite{haimovici2013brain, barzon2022criticality, korchinski2021criticality, das2019critical}. Connecting such emergent properties with information and its propagation is a promising avenue for future works. Overall, our framework makes an important step towards unveiling how information emerges from the structure of multilayer networks, providing fundamental insights into the connection between function, dynamics, and topology. We believe that our work and its future extensions will pave the way for the study of regulatory and control mechanisms in a variety of physical systems, where their multiscale nature shapes their emergent properties.

\begin{acknowledgments}
\noindent G.N. acknowledges funding provided by the Swiss National Science Foundation through its Grant CRSII5\_186422.
\end{acknowledgments}

\appendix
\section{Explicit solution for Gaussian processes on two-layer networks}
\label{app:solution_gaussian}
\noindent Here we present a detailed solution of a general Gaussian process on a multilayer network with two layers and an arbitrary topology. We do not assume any relation between the timescales a priori and show how to recover the timescale-separated solution in suitable limits. Without loss of generality, we start with the case of zero mean $\hat{\theta}$ for simplicity.

To shed light on the role of direct and feedback interactions across layers, we consider that the first layer influences the second, but not vice versa. The transpose adjacency tensor of the multilayer network is
\begin{equation}
    \hat{A} = \begin{pmatrix}
        \hat{A}_{11} & 0 \\
        \hat{A}_{21} & \hat{A}_{22}
    \end{pmatrix}
\end{equation}
so that $\hat{A}_{\mu\mu}$ is the adjacency matrix of layer $\mu$, and $\hat{A}_{21}$ contains the connections from nodes of the first layer to nodes of the second. As in the main text, the first layer has $M_1$ nodes, and the second $M_2$. We do not assume any topology for $\hat{A}$, so long as its spectral radius is such that the dynamics is stable. The multilayer Fokker-Planck operator of the whole Gaussian process is given by
\begin{equation}
    \mathcal{L} = \frac{1}{\tau_1}\mathcal{L}_1 + \frac{1}{\tau_2}\mathcal{L}_2
\end{equation}
with
\begin{align}
    \mathcal{L}_1 & = \nabla_1 \cdot \left[- \vec{x}_1 + \hat{A}_{11} \vec{x}_1 + \hat{D}_1 \nabla_1\right] \nonumber \\
    \mathcal{L}_2 & = \nabla_2 \cdot \left[- \vec{x}_2 + \hat{A}_{22} \vec{x}_2 + \hat{A}_{21} \vec{x}_1 + \hat{D}_2 \nabla_2\right]
\end{align}
where $\vec{x}_\mu$ is the state of each node in layer $\mu$, and $\hat{D}_\mu$ is the diffusion matrix of the layer. The stationary probability of this process is a Gaussian distribution $p_{12} \sim \mathcal{N}(0, \hat{\Sigma})$, where the covariance matrix $\hat\Sigma$ obeys the Lyapunov equation \cite{gardiner}
\begin{equation}
    (\hat{W}\odot \hat{T})\hat{\Sigma} + \hat{\Sigma} (\hat{W}\odot \hat{T})^T = 2 \hat{D}
\end{equation}
where $\hat{D} = \mathrm{diag}(\hat{D}_1/\tau_1, \hat{D}_2/\tau_2)$ is a block-diagonal matrix, and
\begin{align*}
    (\hat{W}\odot \hat{T}) & = \begin{pmatrix}
        \hat{W}_{11} & 0 \\
        \hat{A}_{21} & \hat{W}_{22}
    \end{pmatrix} \odot \begin{pmatrix}
        1/\tau_1 & 1/\tau_1 \\
        1/\tau_2 & 1/\tau_2
    \end{pmatrix} \\
    & = \begin{pmatrix}
        \hat{W}_{11}/\tau_1 & 0 \\
        \hat{A}_{21}/\tau_2 & \hat{W}_{22}/\tau_2
    \end{pmatrix} \\
    & = \begin{pmatrix}
        (\hat{A}_{11} - \mathbb{I})/\tau_1 & 0 \\
        \hat{A}_{21}/\tau_2 & (\hat{A}_{22} - \mathbb{I})/\tau_2
    \end{pmatrix}
\end{align*}
with $\mathbb{I}$ the identity matrix. We can exploit the block structure of the problem to rewrite the Lyapunov equation as the set of equations:
\begin{equation}
\label{eqn:app:Lyap_blocks}
    \begin{cases}
        \hat{W}_{11}\hat{\Sigma}_{11} + \hat{\Sigma}_{11} \hat{W}_{11}^T = 2 \hat{D_1} \\
        \tau_2\hat{W}_{11}\hat{\Sigma}_{12} = - \tau_1 \hat\Sigma_{11}\hat{A}_{21}^T - \tau_1 \hat\Sigma_{12}\hat{W}_{22}^T \\
        \hat{A}_{21}\hat\Sigma_{12} + \hat\Sigma_{12}\hat{A}_{21}^T + \hat{W}_{22}\hat{\Sigma}_{22} + \hat{\Sigma}_{22} \hat{W}_{22}^T = 2 \hat{D_2}
    \end{cases}
\end{equation}
where we $\hat{\Sigma}_{11}$, $\hat{\Sigma}_{22}$, and $\hat{\Sigma}_{12}$ are, respectively, the covariance matrix of the first layer, of the second layer, and across the two layers. In particular, the mutual information between the two layers, $I_{12}$, is given by
\begin{equation}
    I_{12} = \frac{1}{2}\log \frac{\det\hat{\Sigma}_{11}\det\hat{\Sigma}_{22}}{\det\hat{\Sigma}}
\end{equation}
and thus is fully specified by the non-trivial solution of Eq.~\eqref{eqn:app:Lyap_blocks}. Notice that the two timescales of the layers manifestly enter both in $\hat{\Sigma}_{12}$ and $\hat{\Sigma}$.

We now consider the two timescale separation limits. If $\tau_1 \gg \tau_2$, $\hat{A}_{21}$ describes feedback interactions from the nodes of the slow layer, $1$, to those of the fast layer, $2$. In this case, our decomposition of the probability distribution gives (Eq.~\eqref{eqn:multilayer_probability}):
\begin{equation}
    p^\mathrm{st}_{12} = p^\mathrm{st}_{1}p^\mathrm{st}_{2|1} = \mathcal{N}_1(0, \hat{S}_{1}) \, \mathcal{N}_2(\vec{m}(\vec{x}_1), \hat{S}_{2})
\end{equation}
where $\vec{m}(\vec{x}_1) = - \hat{W}_{22}^{-1}\hat{A}_{21} \vec{x}_1$, and the covariance matrices obey the Lyapunov equations
\begin{align*}
        \hat{W}_{11}\hat{S}_{1} + \hat{S}_{1} \hat{W}_{11}^T & = 2 \hat{D_1} \\
        \hat{W}_{22}\hat{S}_{2} + \hat{S}_{2} \hat{W}_{22}^T & = 2 \hat{D_2}  \; .
\end{align*}
Notice that, in the presence of a non-zero $\vec\theta$, the mean of the distribution of the layer $\mu$ is simply shifted by $\theta_\mu$, and the interaction from layer $1$ to $2$ will have $(\vec{x}_1 - \vec\theta_1)$ in the term $\vec{m}(\vec{x}_1)$. These effective probabilities are the stationary solutions of the two effective Fokker-Planck operators 
\begin{align}
    \mathcal{L}_1^\mathrm{eff} & = \nabla_1 \cdot \left[- \vec{x}_1 + \hat{A}_{11} \vec{x}_1 + \hat{D}_1 \nabla_1\right] \nonumber \\
    \mathcal{L}_{2|1}^\mathrm{eff} & = \nabla_2 \cdot \left[- \vec{x}_2 + \hat{A}_{22} (\vec{x}_2 + \hat{A}_{22}^{-1}\hat{A}_{21} \vec{x}_1) + \hat{D}_2 \nabla_2\right] \; .
\end{align}
We can rewrite $p_{12}$ as the Gaussian distribution
\begin{equation*}
    p^\mathrm{st}_{12}(\vec{x}_1, \vec{x}_2) = \frac{1}{\sqrt{(2\pi)^{M_1 + M_2}\det \hat{S}}} \exp\left[-\frac{1}{2} \vec{x}^T \hat{S}^{-1} \vec{x}\right]
\end{equation*}
where the inverse of the covariance matrix $\hat{S}^{-1}$ is
\begin{equation}
\hat{S}^{-1} = 
    \begin{pmatrix}
        \hat{S}_{1}^{-1} + \hat{Z}\hat{W}_{22}^{-1}\hat{A}_{21} & \hat{Z} \\
        \hat{Z}^T & \hat{S}_{2}^{-1}
    \end{pmatrix}
\end{equation}
and $\hat{Z} = \hat{A}_{21}^T (\hat{W}_{22}^{-1})^T \hat{\Sigma}_{22}^{-1}$. We can explicitly invert this expression, finding
\begin{equation*}
    \hat{S} = 
    \begin{pmatrix}
        \hat{S}_{1} & -\hat{S}_{1}\hat{A}_{21}^T (\hat{W}_{22}^{-1})^T \\
        -\hat{W}_{22}^{-1}\hat{A}_{21}\hat{S}_{1} & \hat{S}_{2} + \hat{W}_{22}^{-1} \hat{A}_{21} \hat{S}_{1} \hat{A}_{21}^T (\hat{W}_{22}^{-1})^T
    \end{pmatrix} \; .
\end{equation*}
Clearly, in this limit, the two layers are not independent. The mutual information is given by
\begin{equation}
    I_{12} = \frac{1}{2}\log \frac{\det \left(\hat{S}_{2} + \hat{W}_{22}^{-1} \hat{A}_{21} \hat{S}_{1} \hat{A}_{21}^T (\hat{W}_{22}^{-1})^T\right)}{\det \hat{S}_{2}} \; .
\end{equation}
We remark that it is easy to check that this solution is the same as the one obtained from the full system, Eq.\eqref{eqn:app:Lyap_blocks}, in the limit $\tau_1 \gg \tau_2$: 
\begin{equation*}
    \begin{gathered}
        \hat{W}_{11}\hat{\Sigma}_{11} + \hat{\Sigma}_{11} \hat{W}_{11}^T = 2 \hat{D_1} \\
        \hat\Sigma_{12} = - \hat\Sigma_{11}\hat{A}_{21}^T(\hat{W}^{-1}_{22})^T \\
        \hat{A}_{21}\hat\Sigma_{12} + \hat\Sigma_{12}\hat{A}_{21}^T + \hat{W}_{22}\hat{\Sigma}_{22} + \hat{\Sigma}_{22} \hat{W}_{22}^T = 2 \hat{D_2}
    \end{gathered}
\end{equation*}
which immediately shows that $\hat{S}_1 = \hat\Sigma_{11}$, as expected, while $\hat{S}_2 = \hat{\Sigma}_{22} + \hat{A}_{22}^{-1} \hat{A}_{21} \hat{S}_{1} \hat{A}_{21}^T (\hat{A}_{22}^{-1})^T$.

In the opposite limit, $\tau_1 \ll \tau_2$, the matrix $\hat{A}_{21}$ contains only direct interactions from the nodes of the fast layer, $1$, to those of the slow layer, $2$. Our general decomposition of the probability distribution now reduces to the factorized multilayer distribution
\begin{equation}
\label{eqn:app:p12_direct}
    p^\mathrm{st}_{12} = p^\mathrm{st}_1 p_2^\mathrm{eff, st} = \mathcal{N}_1(0, \hat S_{1}) \, \mathcal{N}_2(0, \hat S_{2}) \; .
\end{equation}
Indeed, it is straightforward to check that the effective operators in this case are
\begin{align}
    \mathcal{L}_1^\mathrm{eff} & = \nabla_1 \cdot \left[- \vec{x}_1 + \hat{A}_{11} \vec{x}_1 + \hat{D}_1 \nabla_1\right] \nonumber \\
    \mathcal{L}_2^\mathrm{eff} & = \nabla_2 \cdot \left[- \vec{x}_2 + \hat{A}_{22} \vec{x}_2 + \hat{D}_2 \nabla_2\right]
\end{align}
from the fact that $\ev{\vec{x}_1}_{p_1} = 0$. In the presence of a non-zero $\vec\theta$, notice that again $\langle \vec{x}_1 - \vec\theta_1 \rangle_{p_1} = 0$, since this difference is the one that enters in the interaction term of the Fokker-Planck operator. Thus, Eq.~\eqref{eqn:app:p12_direct} still holds with both means shifted by $\vec\theta_\mu$. Once more, this solution can be recovered from that of the full system in Eq.~\eqref{eqn:app:Lyap_blocks}, which in this limit gives 
\begin{equation}
    \begin{gathered}
        \hat{W}_{11}\hat{\Sigma}_{11} + \hat{\Sigma}_{11} \hat{W}_{11}^T = 2 \hat{D_1} \\
        \hat{\Sigma}_{12} = 0 \\
        \hat{W}_{22}\hat{\Sigma}_{22} + \hat{\Sigma}_{22} \hat{W}_{22}^T = 2 \hat{D_2}
    \end{gathered}
\end{equation}
and thus $\hat{S}_1 = \hat\Sigma_{11}$ and $\hat{S}_2 = \hat\Sigma_{22}$. Since $p_{12}^\mathrm{st}$ is factorized, the mutual information between the layers is zero.

Thus, in this case of Gaussian processes, we can show explicitly that the separation between the timescale of the two layers modulates the covariance matrix across the layers, creating effective dependencies as the interactions become more and more of the feedback type.

\section{Explicit solution for a network topology with a minimal propagation path}
\label{app:solution_propagation}
\noindent Here we present the explicit solution for the system in Figure \ref{fig:figure2}e, where interactions between layers form the minimal propagation path $\vec{x}_3 \to \vec{x}_1 \to \vec{x}_2$. The adjacency tensor is given by
\begin{equation}
    \hat{A} = \begin{pmatrix}
        \hat{A}_{11} & 0 & \hat{A}_{13} \\
        \hat{A}_{21} & \hat{A}_{22} & 0 \\
        0 & 0 & \hat{A}_{33}
    \end{pmatrix} \;,
\end{equation}
and we assume that the dynamics is described by the Gaussian process in Eq.~\eqref{eqn:langevin_GP} with a non-zero mean.

Following the differential operators defined by the timescale separation solution, Eq.~\eqref{eqn:effective_operators}, we have that the Fokker-Planck operator for the first layer is:
\begin{align}
    \mathcal{L}_{1|3} = - \nabla_1 \biggl[& \hat{W}_{11}\left(\vec{x}_1 - \vec{\theta}_1\right) + \hat{A}_{13}\left(\vec{x}_3 - \vec{\theta}_3\right) + \nonumber \\
    & + \hat{D}_1 \nabla_1\biggl]
\end{align}
since $\rho(1) = \{3\}$, and $\hat{W}_{11} = \hat{A}_{11} - \mathbb{I}$. Hence, we obtain
\begin{align}
    \mathcal{L}_{1|3} = - \nabla_1 \biggl[& \hat{W}_{11}\left(\vec{x}_1 - \vec{m}_1\left(\vec{x}_3\right)\right) + \hat{D}_1 \nabla_1\biggl]
\end{align}
where we introduced the effective mean
\begin{equation}
    \vec{m}_1\left(\vec{x}_3\right) = \vec{\theta}_1 - \hat{W}_{11}^{-1}\hat{A}_{13}\left(\vec{x}_3 - \vec{\theta}_3\right) \;.
\end{equation}
This corresponds to solving the dynamics of the first layer conditioned on the state of the third. The corresponding conditional distribution at stationarity reads
\begin{equation}
    p^\mathrm{st}_{1|3}(\vec{x}_1 | \vec{x}_3) = \mathcal{N}_1(\vec{m}_1(\vec{x}_3), \hat{S}_1)
\end{equation}
where $\hat{W}_{11} \hat{S}_1 + \hat{S}_1\hat{W}_{11}^T = 2 \hat{D}_1$.

We now seek to obtain the effective operator of the second layer. From the definition, Eq.~\eqref{eqn:effective_operators}, since
\begin{equation*}
    \int d\vec{x}_1 \, p_{1 | 3}(\vec{x}_1 | \vec{x}_3) \, \vec{x}_1 = \vec{m}_1(\vec{x}_3)
\end{equation*}
we obtain
\begin{align}
    \mathcal{L}^\mathrm{eff}_{2} = - \nabla_2 \biggl[& \hat{W}_{22}\left(\vec{x}_2 - \vec{m}_2\left(\vec{x}_3\right)\right) + \hat{D}_2 \nabla_2\biggl]
\end{align}
where the effective mean is now
\begin{equation}
    \vec{m}_2\left(\vec{x}_3\right) = \vec{\theta}_2 - \hat{W}^{-1}_{22}\hat{A}_{21}\hat{W}_{11}^{-1}\hat{A}_{13}\left(\vec{x}_3 - \vec{\theta}_3\right) \;.
\end{equation}
Thus, the second layer inherits from the first an explicit dependence on $\vec{x}_3$. The corresponding stationary distribution is
\begin{equation}
    p^\mathrm{eff, st}_{2|3}(\vec{x}_2 | \vec{x}_3) = \mathcal{N}_2(\vec{m}_2(\vec{x}_3), \hat{S}_2)
\end{equation}
where $\hat{W}_{22} \hat{S}_2 + \hat{S}_2\hat{W}_{22}^T = 2 \hat{D}_2$.

Finally, the third layer evolves independently of all others, so that
\begin{equation}
    \mathcal{L}^\mathrm{eff}_{3} = - \nabla_3 \biggl[ \hat{W}_{33}\left(\vec{x}_3 - \vec{\theta}_3\right) + \hat{D}_3 \nabla_3\biggl]
\end{equation}
which implies that $p_3(\vec{x}_3) = \mathcal{N}_3(\vec{\theta}_3, \hat{S}_3)$, and $\hat{W}_{33} \hat{S}_2 + \hat{S}_3\hat{W}_{33}^T = 2 \hat{D}_3$. Thus, the multilayer probability is:
\begin{align*}
    p^\mathrm{eff, st}_{123}(\vec{x}_1, \vec{x}_2, \vec{x}_3) & = \mathcal{N}_1(\vec{m}_1(\vec{x}_3), \hat{S}_1)\, \mathcal{N}_2(\vec{m}_2(\vec{x}_3), \hat{S}_2) \times \\
    & \times \mathcal{N}_3(\vec\theta_3, \hat{S}_3)
\end{align*}
as in the main text. We note that $p^\mathrm{eff, st}_{123}(\vec{x}_1, \vec{x}_2, \vec{x}_3)$ is still a Gaussian distribution. However, our solution makes its conditional structure, and thus the information between the layers, manifest.


%

\end{document}